\documentclass{article}
\usepackage{graphicx}
\usepackage[final]{ap_2021} 
\usepackage[utf8]{inputenc} 
\usepackage[T1]{fontenc}    
\usepackage{hyperref}       
\usepackage{url}            
\usepackage{booktabs}       
\usepackage{amsfonts}       
\usepackage{nicefrac}       
\usepackage{microtype}      
\usepackage{todonotes}

\graphicspath{ {./images/} } 

\title{Alexa, play with robot: Introducing the First Alexa Prize SimBot Challenge on Embodied AI}

\author{
	\And Hangjie Shi
	\And Leslie Ball
	\And Govind Thattai
	\And Desheng Zhang
	\And Lucy Hu
	\And Qiaozi Gao
    \And Suhaila Shakiah
	\And Xiaofeng Gao
	\And Aishwarya Padmakumar
	\And Bofei Yang
	\And Cadence Chung
	\And Dinakar Guthy
	\And Gaurav Sukhatme
	\And Karthika Arumugam
	\And Matthew Wen
	\And Osman Ipek
	\And Patrick Lange
	\And Rohan Khanna
	\And Shreyas Pansare
	\And Vasu Sharma
	\And Chao Zhang
	\And Cris Flagg
	\And Daniel Pressel
	\And Lavina Vaz
	\And Luke Dai
	\And Prasoon Goyal
	\And Sattvik Sahai
	\And Shaohua Liu
	\And Yao Lu
	\And Anna Gottardi
	\And Shui Hu
	\And Yang Liu
	\And Dilek Hakkani-Tur
	\And Kate Bland
	\And Heather Rocker
	\And James Jeun
	\And Yadunandana Rao
	\And Michael Johnston
	\And Akshaya Iyengar
	\And Arindam Mandal
	\And Prem Natarajan
	\And Reza Ghanadan
}

\begin{document}

\maketitle

\begin{abstract}
   The Alexa Prize program has empowered numerous university students to explore, experiment, and showcase their talents in building conversational agents through challenges like the SocialBot Grand Challenge and the TaskBot Challenge. As conversational agents increasingly appear in multimodal and embodied contexts, it is important to explore the affordances of conversational interaction augmented with computer vision and physical embodiment. This paper describes the SimBot Challenge, a new challenge in which university teams compete to build robot assistants that complete tasks in a simulated physical environment. This paper provides an overview of the SimBot Challenge, which included both online and offline challenge phases. We describe the infrastructure and support provided to the teams including Alexa Arena, the simulated environment, and the ML toolkit provided to teams to accelerate their building of vision and language models. We summarize the approaches the participating teams took to overcome research challenges and extract key lessons learned. Finally, we provide analysis of the performance of the competing SimBots during the competition.
\end{abstract}

\section{Introduction}


Conversational assistants such as Amazon Alexa, Apple's Siri, and Google Assistant have become an increasingly commonplace way for people to access information and content and control connected devices such as smart outlets, lighting, and home security systems. A key frontier in conversational AI is to advance from spoken dialog and enable embodied conversational systems where the conversational agent is able to perceive the physical world, navigate within it, and move and manipulate objects. In the future, we envision a world where everyday conversational assistants will be able to navigate and actuate in the physical world. They could, for example, make you an omelette, pour you a cup of coffee, explore your house to find your slippers, or identify and address an issue such as a door left open or a leaking faucet.

The Alexa Prize\footnote{\url{https://www.amazon.science/alexa-prize}} is an Amazon Alexa sponsored program that in recent years has enabled hundreds of university students and faculty to compete in advancing the state-of-the-art in conversational AI. Since 2016, the SocialBot Grand Challenge has hosted a competition among universities from across the world to compete in creating the best \textit{SocialBot}, i.e., an Alexa skill that can engage in extended open-domain dialogs with users around popular topics and current events~\cite{hu2021further}. Since 2021, the TaskBot Challenge has engaged teams in building conversational assistants that can assist users in completing complex tasks such as recipes or Do It Yourself (DIY) projects~\cite{gottardi2022alexa}. One of the key advantages of the program is that it enables university teams to rapidly test and iterate on their approaches through testing with real-world users at scale through Alexa.

\begin{figure}[!h]
\centering
\includegraphics[width=1\textwidth]{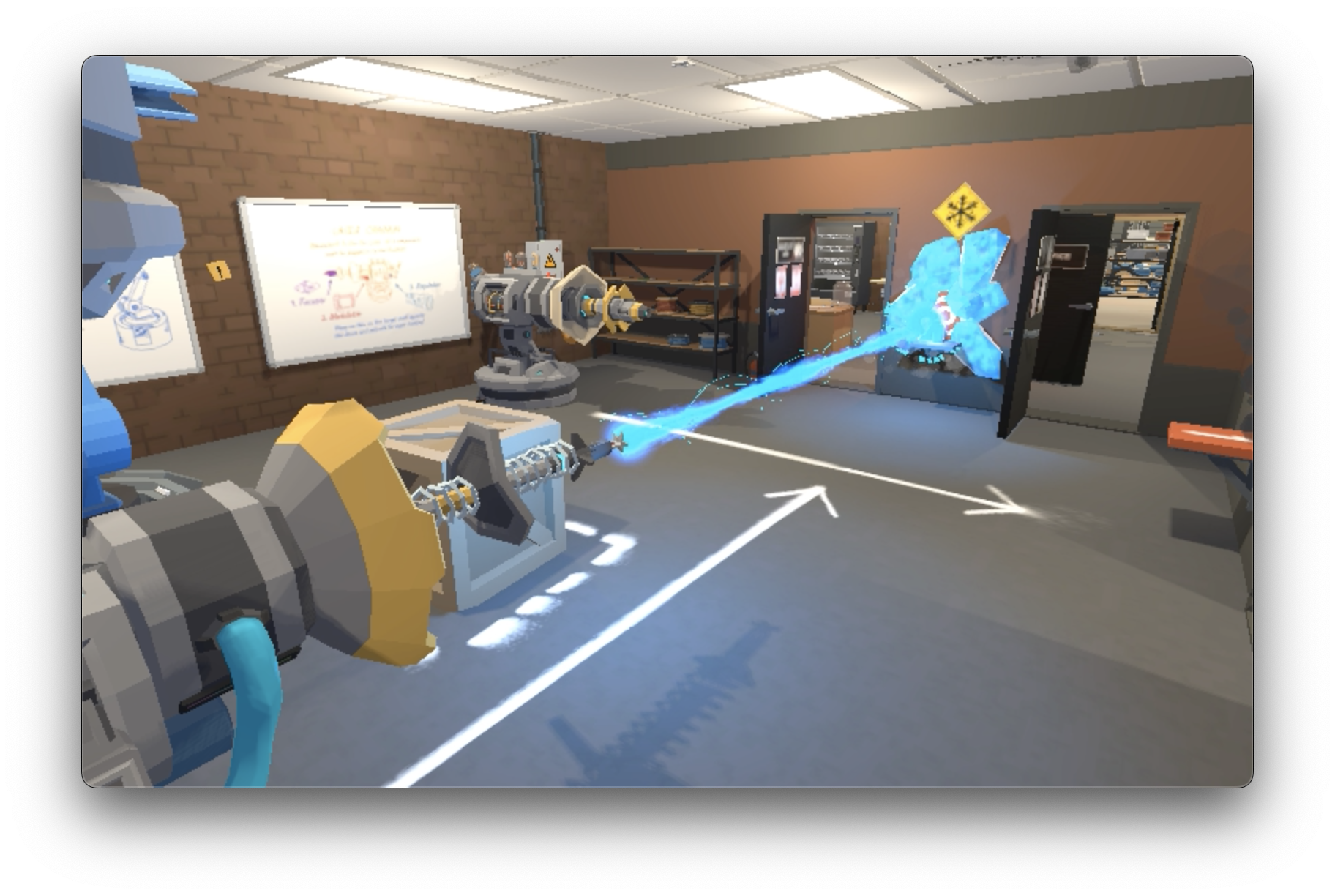}
\caption{ The egocentric view of the robot in a simulated room. }
\label{fig:egocentric_view}
\end{figure}

The motivation for the third Alexa Prize Challenge, the SimBot Challenge, is to push the boundaries of embodied AI and drive towards the vision above. Inspired by the significant role that games have played in showcasing and fostering the evolution of core AI capabilities, the SimBot Challenge incorporates elements of game design into the development process. As Zobrist’s paper ~\cite{zobrist1969model} on the application of AI to the game Go emphasizes, "More study of this complex game may reward us with new insight into human perceptual and problem solving abilities as well as foster the development of new techniques for artificial intelligence." To facilitate interaction with the system by Alexa users and as a precursor to physical embodiment, the SimBot Challenge presents a simulated office/lab environment named Alexa Arena~\cite{gao2023alexa}, created using the Unity gaming engine. This environment comprises multiple rooms, each equipped with various devices and objects. Users interact with the environment by giving instructions to a robot companion, the SimBot. By adopting a gaming-inspired framework, the challenge provides users with an engaging and dynamic environment to interact with the simulation through screen-enabled devices such as Echo Show and FireTV.
The SimBots respond to spoken commands from users, taking actions to move in and manipulate the environment, asking verbal questions, and providing feedback. The fusion of game mechanics and AI development allows for an immersive user experience, bridging the gap between virtual simulation and physical embodiment. In Figure~\ref{fig:egocentric_view}, we can see the robot's view of the simulated environment display on an Echo Show device. The SimBots were launched first for testing with a cohort of Amazon employees in November 2022, followed by a general public launch in December 2022 at which time Alexa users in the United States with supported devices could interact with the participating SimBots by saying "alexa play with robot" to a screen-enabled Alexa device.

As an initial pre-phase of the program, we conducted an offline evaluation using the TEACh embodied AI dataset~\cite{teach}. We summarize this phase in Section~\ref{sec:offline_challenge}. In Section~\ref{sec:online_challenge}, we detail the operation of the online challenge. In Section~\ref{sec:capabilities}, we describe the infrastructure supporting the challenge and the tools and capabilities provided to teams. In Section ~\ref{sec:scientific_partipants}, we discuss the scientific innovations and advancements in the competition. The performance of the SimBots is reviewed in Section~\ref{sec:results} and insights gathered from the SimBot Challenge are discussed in Section~\ref{sec:conclusions}.









\section{Offline Challenge}
\label{sec:offline_challenge}
As an initial phase for teams to develop modeling capabilities for embodied task completion, an offline challenge was conducted using the TEACh dataset~\cite{teach}. The TEACh dataset consists of annotators role playing interactions between a Commander (User) and Follower (Robot) collaborating to complete tasks in a simulated home environment. The data was collected using the AI2-THOR simulator~\cite{ai2thor}, employing a web interface that allowed both annotators to navigate and observe the simulated home from a first-person perspective. In this setup, only the Commander had access to the details of the task, while the Follower could interact with objects in the room to complete the task, necessitating communication and coordination between them through live text chat. To encourage multi-turn interactions, the Commander could additionally search for objects that were not directly visible and provide appropriate directions to the Follower. The Follower could perform 8 possible object interaction actions - \textit{Pickup}, \textit{Place}, \textit{Open}, \textit{Close}, \textit{ToggleOn}, \textit{ToggleOff}, \textit{Slice}, and \textit{Pour}. Successful task completion required navigating around the room, searching for objects inside receptacles such as cabinets or refrigerators, and reasoning about physical state changes (e.g. place a slice of potato on a pan located on the stove and turn on the stove to cook it). In each data collection session, the initial state, dialogue utterances and actions taken by each annotator were saved so that the gameplay session could be replayed in the AI2-THOR environment.

The Alexa Prize SimBot Offline Challenge required teams to build models for the Execution from Dialog History (EDH) task based on this dataset. Given some dialogue history and a partial sequence of actions and corresponding egocentric image observations from the Follower, an EDH model should predict subsequent actions for the Follower. The EDH instances are constructed from data collection sessions by examining the action sequence between every pair of dialogue utterances. The target action sequences are selected based on criteria such as non-empty preceding dialogue history, presence of at least one object interaction within the action sequence, and inclusion of the state changes in at least one task-relevant object.

An EDH model receives input comprising the dialogue history, history of actions by the Follower, and their corresponding egocentric image observations. At each time step, the model is responsible for predicting an action which could be an object interaction, a navigation action, or a special \textit{Stop} action. If the model predicts an object interaction action, it must additionally predict an $(x, y)$ coordinate in the egocentric observation of the Follower to identify the target object for the action. After the action predicted by the model is executed in the simulator, the simulator state is updated and the model receives an updated egocentric image observation. The execution process continues until the model predicts the \textit{Stop} action, 1,000 actions are executed or 30 actions result in API failures. Models are evaluated by comparing the state changes resulting from the models' predicted actions with those taken by the annotators.

Teams were provided with code to replay TEACh sessions as well as wrapper code to train and perform inference for TEACh EDH models. Furthermore, a baseline model based on the Episodic Transformer~\cite{et} model was provided to the teams. To proceed to the next stage of the challenge, teams needed to develop and submit a model that outperformed the baseline ET model.

\section{Online Challenge}
\label{sec:online_challenge}


The next phase of the challenge was an online challenge, where models are integrated into a runtime service to support a real-time gaming experience on Alexa multimodal devices. In the online experience, a robot operates within a simulated office/lab environment powered by the Unity gaming engine. University teams were tasked with setting up a robot action inference service supported by their vision models, which seamlessly integrated into a runtime system developed by the Alexa Prize team. The entire interaction experience is built on top of Alexa skills, leveraging the Alexa Web API for Games interface. Users can engage with the robot through voice commands and observe updates in the simulated environment through video streaming to their device. Voice requests transmitted via the device are converted to text by Alexa speech recognition services, initially processed by the SimBot skill. The user command is then forwarded to the SimBot runtime system, where it undergoes further interpretation to generate executable actions within the simulated environment. The model is responsible for predicting the robot's actions or engaging in dialog based on the input utterance text and the image of the robot's egocentric view. Upon concluding the interaction with the SimBot, users are presented with an opportunity to provide feedback in the form of a verbal rating and optional free-form comments. These ratings and feedback are valuable resources shared with the university teams, enabling them to gain insights and make improvements to their model performance.

The SimBot online phase began with a comprehensive three-day Bootcamp in August 2022. During this event, ten university teams were exclusively invited to receive Amazon Web Service (AWS) training, SimBot tooling, and hands-on development experience. Throughout the Bootcamp, all ten teams successfully developed a SimBot using a baseline model provided by Alexa Prize, utilizing the resources offered by AWS and Alexa. Following this training, teams put their efforts into refining and enhancing their SimBots until the end of October, ultimately completing the skill certification process required for integration with Alexa users. An initial feedback phase then took place to gather early feedback from beta users, followed by the general availability launch in December 2022. All ten teams progressed from the launch phase and advanced to the Semifinals from February 2, 2023 through March 22, 2023. From the Semifinals, five teams successfully qualified as Finalists and participated in the Finals phase that ended on April 28, 2023. The closed-door Finals event took place on May 3, 2023, where the teams competed for the top honors.

\subsection{Capabilities Provided to Teams}
\label{sec:capabilities}
To facilitate the development of SimBot, the university teams were granted exclusive access to a range of Amazon Alexa resources, technologies, and experts. The following is an overview of the resources that were made available to them.

\subsubsection{Alexa Arena Simulated Environment}

Alexa Arena~\cite{gao2023alexa} is a Unity-based 3D embodied AI simulator built by Amazon Alexa AI. In Alexa Arena, an agent acts in a 3D environment that supports a variety of indoor object interaction tasks. Alexa Arena features high-quality graphics, animations, navigation and object manipulation to enable highly interactive and user-centric multimodal embodied AI research.

There are 336 unique objects in Alexa Arena. Each object has a set of properties (i.e., affordances), which specify if a certain type of robot-object interaction is possible. For example, the agent can toggle the \textit{3-D printer} since it has an object property \textit{toggleable}. In total, there are 14 object properties, including \textit{pickupable}, \textit{openable}, \textit{breakable}, \textit{receptacle}, \textit{toggleable}, \textit{powerable}, \textit{dirtyable}, \textit{heatable}, \textit{eatable}, \textit{chillable}, \textit{fillable}, \textit{cookable}, \textit{infectable}, and \textit{decor}. Each object property has a corresponding action and object state to go into when acted on. For example, \textit{break} is the corresponding action for \textit{breakable}, and \textit{broken} is the corresponding state after the action has been performed.

In Alexa Arena, robot actions can be categorized into two types: 1) user interaction actions for communicating with the user via starting a dialog or highlighting objects in the scene\footnote{Note that highlighting is used as proxy for deictic gestures by the robot. The current generation of SimBots are  not able to point using their arms.}, and 2) robot physical actions to interact with the simulation environment. Robot physical actions include both navigation and object interaction. To improve the user experience, all the navigation and interaction actions are animated in a continuous fashion and accompanied by environmental sounds.

\subsubsection{ML Toolkit}

\begin{figure}[!h]
\centering
\includegraphics[width=1\textwidth]{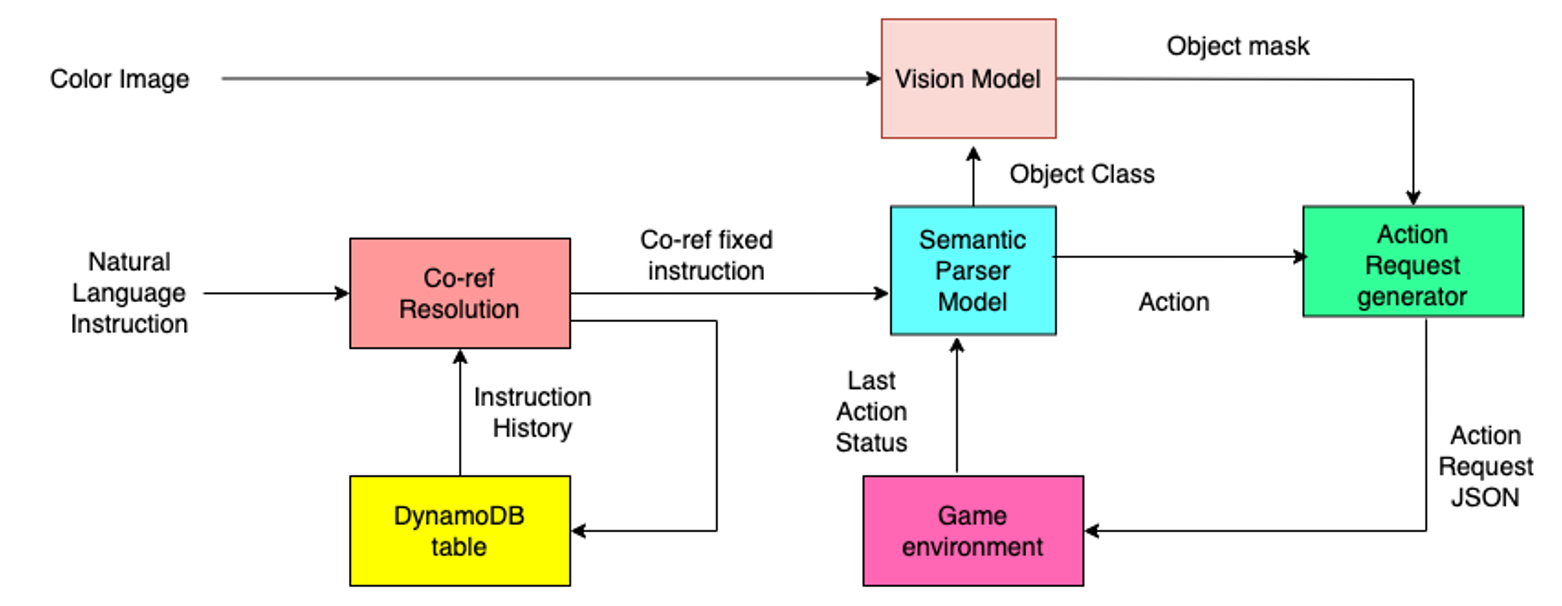}
\caption{ The provided baseline model in ML Toolkit. }
\label{fig:placeholder_model}
\end{figure}

Along with the Alexa Arena simulator, we also provided teams with an ML toolkit to support model development. This toolkit provides a baseline robot model (Figure~\ref{fig:placeholder_model}) that can handle basic visual perception, action prediction, and dialog management for completing game missions in the SimBot Challenge. Furthermore, the toolkit includes two datasets to aid in the training of robot models. The first dataset is a hybrid dataset where ground-truth robot action trajectories are paired with human annotated dialogs. The second dataset comprises a collection of over 600,000 images labeled with object segmentation, which can be used to train object detection models.

Within the baseline model, the visual perception module is a Mask-RCNN model trained on the provided image dataset. This model takes an RGB image as input and predicts masks for all object instances (across 86 object classes) present in the image. This model exhibits reasonable object detection performance on the validation set. Table~\ref{tab:vision_results} shows its mean Average Precision (mAP) for small, medium and large objects.

\begin{table}[h]
\centering
\begin{tabular}{|l|c|c|}
    \hline
    \textbf{Obj Category} & \textbf{Area ($px^{2}$)} & \textbf{mAP} \\ \hline
    Small & 0 - 1296 & 37.63  \\ 
    Medium & 1296 - 9216 &60.41 \\  
    Large & 9216 - 90000 & 64.72 \\  
    \hline
    Overall & 0 - 90000 & 46.03  \\ 
    \hline
\end{tabular}
\vspace{2mm}
\caption{Evaluation results for the provided Mask-RCNN model on small, medium, and large objects.}
\label{tab:vision_results}
\end{table}


\subsubsection{Automatic Speech Recognition and Text to Speech}
To improve the experience for users interacting with SimBots, we supplied Automatic Speech Recognition (ASR) technology that converts user utterances to text and Text-To-Speech (TTS) technology that generates spoken responses from SimBots. Additionally, the participating university teams were given access to tokenized N-best ASR hypotheses that included confidence scores for each token. This resource allowed the teams to fine-tune and optimize their SimBots for more accurate and effective communication with users.  

To further improve the accuracy of ASR, we extended the SimBot skill interaction model and introduced hints for the SimBot skill intents. This model included over 10,000 sample utterances encompassing a diverse range of supported robot actions and objects, and was provided to the teams as a template to incorporate into their models. With a comprehensive set of hints that covered a wide range of possible interactions, the teams could create more accurate models that could better understand and respond to user requests. 

\subsubsection{Infrastructure and SDK}

\begin{figure}[!h]
\centering
\includegraphics[width=1\textwidth]{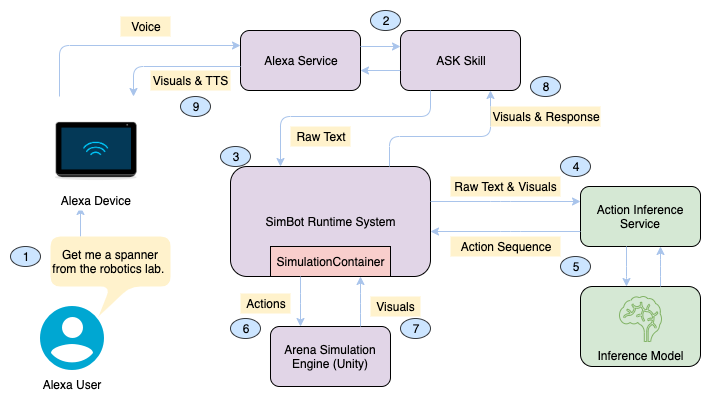}
\caption{ Flow diagram illustrating the sequence flow of an user interaction with SimBot runtime services. }
\label{fig:interaction-sequence-fig}
\end{figure}

As part of the Alexa Prize SimBot Challenge, we provided participating university teams with a powerful runtime system that seamlessly integrates their models into a gaming experience for end-users on various devices, including Echo Show and Fire TV. This system provides an opportunity for teams to showcase their models and offer users an engaging interactive experience. The following sequence flow illustrates the respective steps shown in Figure~\ref{fig:interaction-sequence-fig}, for one interaction with the end-user:
\begin{itemize}
    \item \textbf{1}: Alexa-user interacts with SimBot Skill using natural language instruction, such as "Get me a spanner from the robotics lab".
    \item \textbf{2}: SimBot Skill receives the user utterance, and invokes the Runtime System with the contextual information.
    \item \textbf{3}: SimBot Runtime System sends the image from the robot’s current viewpoint (egocentric view), along with the user’s input utterance, to the Action Inference Service (developed by the respective university team).
    \item \textbf{4-5}: University model predicts the next sequence of actions (e.g. move forward 2 steps), or generates a suitable text response.
    \item \textbf{6-7}: Each of the actions in the predicted sequence (in Step 4) are executed in Arena Simulation Engine (built using Unity), and the visuals are live-streamed to the Alexa device. Note: Steps 4-7 are repeated, to execute subsequent sequences (\textit{look down\textrightarrow find lamp\textrightarrow  toggle on/off}), until the university model determines that the action sequence is complete, and/or generates a text response.
    \item \textbf{8-9}: The language response from the university SimBot (if any) is played on the Alexa device, and the microphone is opened for subsequent interaction with the user.
\end{itemize}

At the end of each turn, the SimBot Runtime System checks the state of the simulation environment, to verify if the respective goal has been completed successfully. Steps [1-9] are repeated until the successful completion of the goal. A user-satisfaction score (1-5) is requested at the end of a game session.

\begin{figure}[!h]
\centering
\includegraphics[width=1\textwidth]{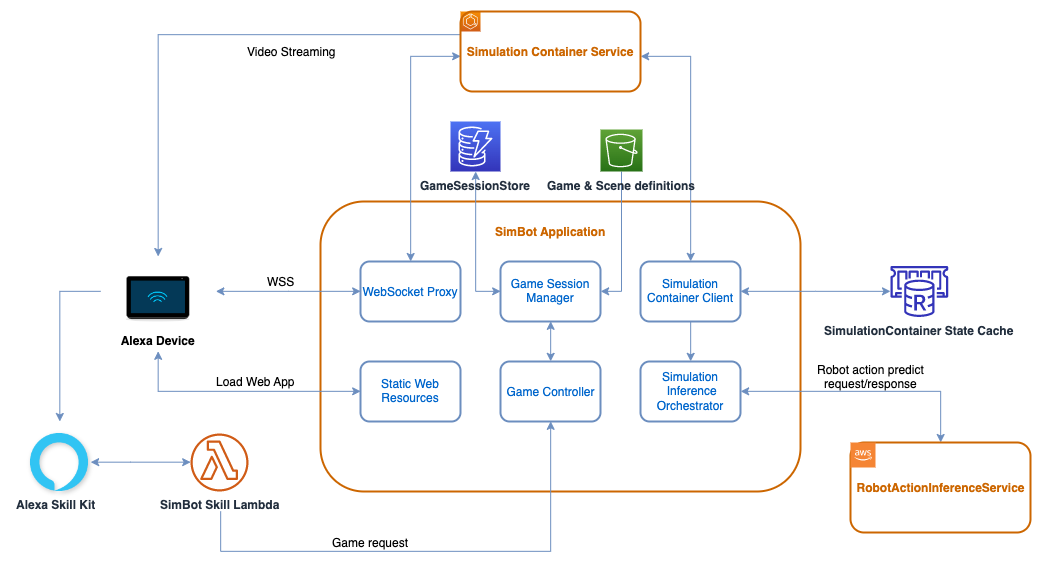}
\caption{ SimBot Runtime System Diagram and Workflow }
\label{fig:system-workflow-fig}
\end{figure}

The sequence flow above involved the main components listed below:
\begin{enumerate}
    \item \textbf{SimBot Application}: A RESTful web application hosted in AWS Elastic Container Service (ECS) Fargate. It manages the lifecycle of a game session with the end user. SimBot application orchestrates the inputs between the robot action inference service and the simulation container. User commands are interpreted into robot actions that can be executed in the simulation container. It also functions as a proxy layer which validates and proxies the WebSocket connection from the multimodal device to the simulation container.
    \item \textbf{Simulation Container Service}: This service acts as a wrapper to the Alexa Arena simulation engine to execute robotic actions and to stream visuals from the engine to upstream applications.
    \item \textbf{Robot Action Inference Service}: This component represents the brain of the embodied agent. Natural language instructions from end-users along with the live visuals from the Alexa Arena simulation engine are processed by the Action Inference service, to generate the respective sequence of robot actions and optional clarification dialog. To achieve this, this service hosts the ML models and performs inferencing at runtime.
    \item \textbf{SimBot Skill}: This is an Alexa skill built on top of the Alexa Skills Kit (ASK). The SimBot skill receives ASR outputs from the Alexa service. An AWS Lambda function handles the skill requests and delegates the requests to the SimBot Application. The skill also uses the Alexa Web API for Games interface which supports the multimodal gaming experiences that run on Alexa-enabled devices.
\end{enumerate}

The Alexa Prize SimBot Challenge provides an opportunity for university teams to prove their expertise in machine learning and conversational AI. To enable university teams to focus on scientific innovation, we developed an SDK that provides a CLI, scripts, and utilities to simplify engineering work and operational tasks. The university teams could spend minimal manual effort executing the SDK, making minor code changes to their own systems and operationally maintaining them once spawned. 

\begin{figure}[!h]
\centering
\includegraphics[width=1\textwidth]{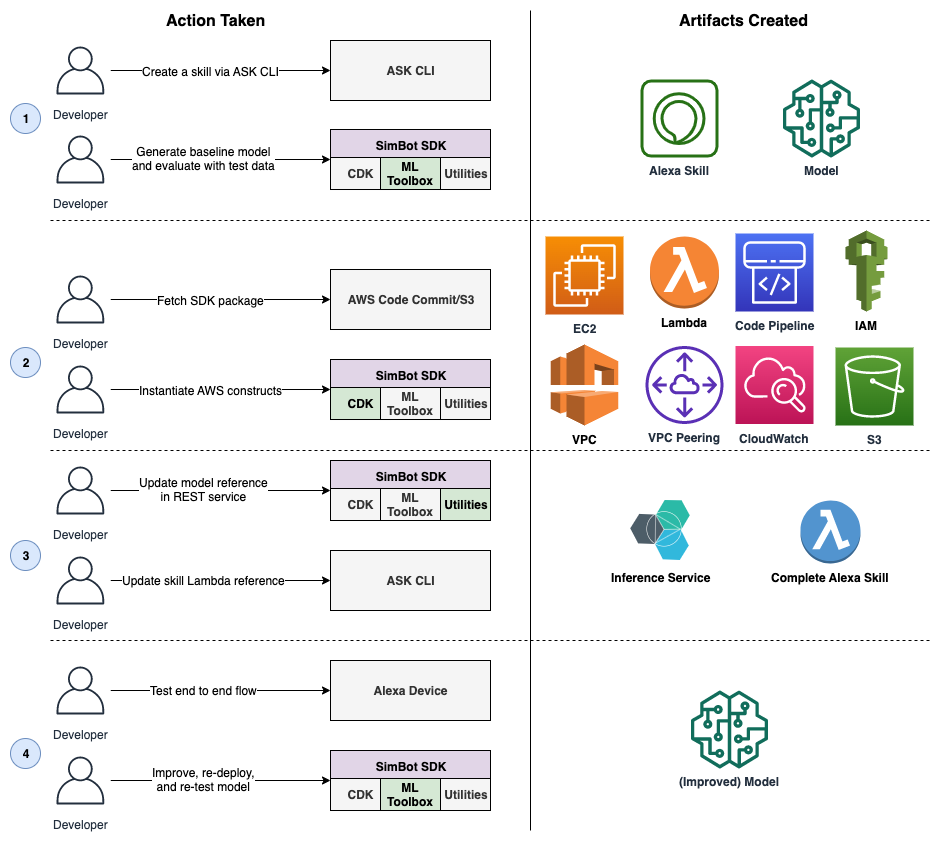}
\caption{ Developer experience for a university team }
\label{fig:sdk-developer-workflow-fig}
\end{figure}

The SimBot SDK leverages the AWS Cloud Development Kit (CDK) to provision and manage resources within their AWS accounts. The CDK application for SimBot automates the deployment of the ASK skill Lambda, Virtual Private Cloud (VPC), Identity Access Management (IAM) role, Cloud Watch logs, and other resources required for the challenge. It provides continuous integration for the Action Inference service and skill Lambda, making it easier for developers to iteratively update the service. It also enforces a separation of test and production stages for enhanced reliability. In addition, the SimBot SDK includes several utilities, such as template implementations based on the API contract of the Action Inference service, integrated with baseline bootstrapped models and DynamoDB tables. The utilities also provide pointers to third-party libraries for ML utilities such as profanity-filter, spaCy, and AllenNLP libraries.

\subsection{Customer Feedback Data and Evaluation Metrics}

A key benefit provided to the university teams was the ability to field their SimBots with Alexa users. Throughout general availability and Semi-finals phases, users interacted with the SimBots and were subsequently prompted for satisfaction ratings and free-form feedback on their experience. In addition, the system was instrumented to capture the duration of conversations, and the status of game mission completion. Mission completion is measured by a metric called Mission Success Rate (MSR), which calculates a team's average rate of successfully completing mission goals in games:
$$
MSR = \frac{N(\textup{succeeded missions})}{N(\textup{total missions})}
$$
The average user satisfaction ratings together with mission success rate served as the primary evaluation metrics for the challenge. Each university team had access to these metrics and also received an anonymized leaderboard daily that presented the average metrics and rankings for all SimBots participating in the challenge. These provided the teams with valuable information to assess their performance and allowed them to gauge their relative performance compared to other teams. In addition, teams had access to transcriptions of the free-form qualitative feedback shared by users at the end of their interactions with the team’s SimBot allowing the teams to gain qualitative insights into the users' impressions of the SimBots.

\subsection{Support from the Alexa Prize team}

In addition to providing data and infrastructure, we engaged with university teams in several ways to provide support and feedback:
\begin{itemize}
    \item A virtual pre-bootcamp to onboard university teams to the SDK and prepare teams for the bootcamp.
    \item A hands-on bootcamp with training materials, best practices, and design guidelines.
    \item Two virtual sessions with university teams on CX design, model training and evaluation, and competition guidelines to prepare teams for each phase of the competition.
    \item An internal beta phase, to provide traffic from Amazon employees to help inform and improve SimBot performance before general availability to all Alexa users.
    \item Detailed report on SimBot experiences prior to public launch, evaluating functionality as well as the SimBot's ability to maintain anonymity and handle inappropriate interactions.
    \item Weekly office hours for 1:1 consultations with a dedicated Alexa Prize Solutions Architect, Program Managers, UX Designers, and members of Alexa science and engineering teams.
    \item On-demand access to Alexa Prize personnel via Slack and email.
\end{itemize}

\section{Scientific Advancements}
\label{sec:scientific_partipants}

During the challenge, the participants worked actively to improve their robots to enhance user satisfaction during game interaction and improve task completion rates. These include scientific innovations and engineering optimizations across multiple areas including data generation and annotation, efficient data storage and retrieval, user interaction tracking, visualization systems, dialog management, visual grounding, action prediction, multimodal and language understanding, and continuous improvement workflows. In this section, we present a summary of the main scientific advancements explored by the participants during the implementation of their robots. Each participating team described their specific innovations in more detail as part of their paper in these proceedings. The scientific contributions span multiple aspects that are instrumental to the seamless functionality of embodied AI agents. End-users interact with the embodied AI agents using voice commands through their Echo Show or Fire TV devices. The voice commands are then transcribed to text using the Alexa ASR system. Following this transcription, teams work with the text input to perform natural language understanding, the first party view of the robot to perform vision understanding and grounding, and combine both modalities to eventually execute the user-intended instruction on Alexa Arena.

Generalizability of models was key scientific theme and influenced the structure of the competition phases. Throughout the challenge, participating robots were evaluated on both seen game missions and unseen game missions. In phases with the seen game missions, teams had the opportunity to play with the games, review user feedback, and update their robots accordingly. During phases with unseen games, the robots were evaluated on their ability to solve missions that had not been seen before, and no updates or modifications to the robot models were permitted. In those unseen games, the robots may encounter new objects and new action types while completing previously unseen mission goals. To tackle these challenges, the teams focused on improving the generalizability of various aspects within their robots, via building (a) robust vision modules that cover all task-related objects, (b) natural language understanding mechanisms that can reliably predict user intents and map them to robot actions, and (c) adaptive dialog management strategies that offer informative responses and valuable guidance to users, even in unseen scenarios.



\subsection{Natural Language Understanding and Action Prediction}
\label{nlu_teams}

During each game mission in the SimBot Challenge, the users are presented with a task description and a list of subgoals, while the SimBot can only get access to this information through the user's language inputs, and in this real world scenario, users can instruct the SimBot in any way they want. The user utterances are often incomplete, ambiguous or completely out-of-domain. Furthermore, user utterances can have different levels of abstraction. Some users may prefer to provide procedural step-by-step instructions (e.g., ``pick up the mug''), while others may prefer to give high-level commands (e.g., ``repair the broken bowl'') or combinations of actions (e.g. ``go to the fridge in the break room and pick up the mug''). This diversity in user instructions poses a major challenge for robustness in language understanding.

To robustly handle user utterances, most teams have adopted modular architectures, where input language is first processed by natural language processing modules (e.g., part of speech tagging, semantic role labeling, named entity recognition, intent classification) or neural models (e.g., Transformer~\cite{vaswani2017attention} based deep models) for identifying the user intent and related objects. The intermediate representation of the input is then mapped to a sequence of robot actions via symbolic planners, pre-defined templates/rules or trained neural models, which often take into consideration the robot state and environment state to make context-aware predictions. Moreover, some teams have also injected common sense knowledge into their action prediction process. For example, knowledge of the affordances of an object can often help the robot eliminate unlikely action-object predictions. One of the major challenges the teams faced was grounding the understanding onto Alexa Arena - teams have proposed a combination of rule-based and neural architectures to do this transformation, making their solutions more versatile to other applications as well. Team EMMA proposed a foundational transformer based end-to-end model with pre-training strategies, datasets and a curriculum of pre-training tasks to train a foundational model before eventually fine-tuning the model for the embodied AI task on Alexa Arena. This approach showed good performance both offline and online. The same team also shows preliminary results on sim-2-real transfer using the proposed pre-training strategy and the provided Alexa Arena datasets.

\subsection{Visual Grounding}
\label{response_teams}

In the game, the robot can observe the 3D scene via a first-party RGBD camera. Any object-related actions (like navigate to an object or manipulate an object) require the robot to provide the correct object mask based on the image from it's current first-party view. Therefore, it is essential for a robot to efficiently recognize objects in the scene and correctly ground user utterances to the corresponding objects.

To ground user instructions to objects in the scene, teams use neural methods to perform object detection (or semantic segmentation). Their contributions involve fine-tuning the baseline mask RCNN model for mask prediction, and building additional models to detect object categories, states and relations. For example, team GauchoAI fine-tuned a MaskFormer model~\cite{cheng2021per} using the provided image dataset, showing better visual understanding capability (absolute improvement of 12\% - 22\% mAP for medium and large objects compared to the provided baseline system). Team Seagull built a hierarchical visual perception system, including a Mask2Former model to detect coarse object types, a ResNet model to detect fine-grained object types and states, and a heuristic method to verify object spatial relations with high accuracy. Team EMMA fine-tuned a pre-trained VinVL model~\cite{zhang2021vinvl} with the Alexa Arena dataset to improve detection accuracy. The numbers are not directly comparable to the baseline model metrics because the team has also modified the number of object detection classes. Additionally, team EMMA also showed preliminary results for sim2real transfer for object detection by benchmarking on a synthetic dataset curated from the public GQA dataset~\cite{hudson2019gqa} showing similar performance among Alexa Arena objects as well as other objects not present in the Alexa Arena dataset. The same team has also trained a visual ambiguity detector module to efficiently ground instructions in cases where there are multiple occurrences of the referred object. The output is modeled as a sequence that first predicts the presence of ambiguity in grounding, which is then used by a downstream grounding module. Team KnowledgeBot used the baseline model to produce object masks but determines which masks to retrieve based on objects generated from their planner. Team SlugJARVIS trained a MaskFormer and a ResNet based classifier model to do both coarse and fine-grained object detection, showing a high accuracy of 93\% on fine-grained object classification. They also coupled an object state detector with an object relation detector to identify object states and spatial relationships between them. Across the teams, visual grounding is performed using heuristics, or through efficient integration of vision language models. Since visual grounding relies on language input, teams have proposed highly interlinked language and visual grounding modules. 

A common challenge in object grounding arises when there are multiple instances of the same object type. Sometimes there are some details in user utterances that can help to disambiguate, for example, location information or object attributes (e.g. color). Most teams have built object attribute detectors based on simple rules or statistical models (e.g. K-means clustering). 

To facilitate efficient navigation and object localization, several teams maintain a symbolic scene representation, including semantic maps and scene graphs, from visual observations at each time step. The representation enables the robot to efficiently explore the virtual environment and navigate to the requested objects. Some teams also introduce a memory bank to incorporate visual memory, which is populated with beliefs about various objects in different rooms that are updated periodically during missions. This approach provides a probability distribution of seen objects for a given location which can be used by the robot for easy navigation when user instructions allude to previously seen objects. 

\subsection{Knowledge}
\label{knowledge_teams}

To efficiently assist users in completing game missions, it is important for the robot to possess enough background knowledge on the mechanism of environment state transition, for example, regarding objects and their properties, actions and their effects. Most teams maintain a collection of offline knowledge sources, including knowledge based on game missions like common action sequences, as well as more general common knowledge like object affordances and object aliases. The offline knowledge provides guidance for action prediction, visual grounding, object disambiguation, dialog management and response generation.

In addition, team SlugJARVIS also maintains and actively updates a set of online knowledge, which contains multimodal information from vision, text, and executed actions. They propose a progressive and evolving task experience retrieval algorithm that can identify unseen tasks and adapt to various environments and tasks by leveraging past successful interactions. 

\subsection{Dialog Management}
\label{dm_teams}

For regular users, the experience of instructing a robot via language to complete game missions is quite different from playing the game mission by themselves, especially when they are not familiar with the robot's capabilities or the limitations of the game environment. Therefore, actively providing appropriate feedback becomes critical for building user trust and delivering an engaging user experience. Most teams propose a multitude of template-based dialog generation modules that are easily extendable and facilitate continuous development. These modules include data structures that store dialog acts, template based dialog generation and tracking, as well as question answering based architectures for understanding user responses. To make the generated responses more natural and human-like, teams also use a variety of techniques including using large language models (LLM) to generate diverse response templates, and adding emotional prosody to the speech. 

To further simplify users' efforts in completing game missions, several teams propose strategies for proactively suggesting next actions based on the current game state. Note that the robot cannot directly access the game mission description; it has to infer the next proper actions based on the dialog history and previous executed actions. For example, team GauchoAI makes suggestions based on objects recently interacted with and their affordance, e.g., when the robot approaches the microwave with a heatable object in hand, it is likely the user wants to heat the object.

\subsection{Training and Data Generation}
\label{training_teams}

Utilizing the provided Alexa Arena simulator, baseline model, and trajectory and vision datasets, several teams have managed to generate more synthetic data to further enhance their model training. These include multimodal vision and language datasets as well as language-based task decomposition and coreference resolution datasets. For example, team ScottyBot uses template-based synthetic language-actions data to train a BART model \cite{lewis2019bart} for action sequence prediction from user utterances. To handle ASR errors, team SlugJARVIS employs LLMs to generate user utterances with simulated ASR errors for action prediction. Other examples include generating multimodal vision and language data, as well as language based coreference resolution data. In addition to generating these datasets, teams build dedicated annotation systems to create and refine these datasets either using offline approaches or by leveraging online user interaction data.


\section{SimBot Performance: Results and Analysis} 
\label{sec:results}


Building on the success of the SocialBot and TaskBot challenges, the users' explicit ratings and feedback were used to evaluate the SimBots. Additionally, a task-oriented measure known as the mission success rate was introduced, allowing for a direct way to evaluate the SimBots' effectiveness in accomplishing the tasks within a game mission. Furthermore, new unseen game missions were introduced during the competition to evaluate the SimBots' generalizability. In this section, we provide various metrics to evaluate the performance of the SimBots in the first year of the competition, including comparisons between the Finalists, all SimBots, and our baseline system.

\subsection{Satisfaction Ratings}
\label{sec:ratings}

\begin{figure}[!h]
\centering
\includegraphics[width=1\textwidth]{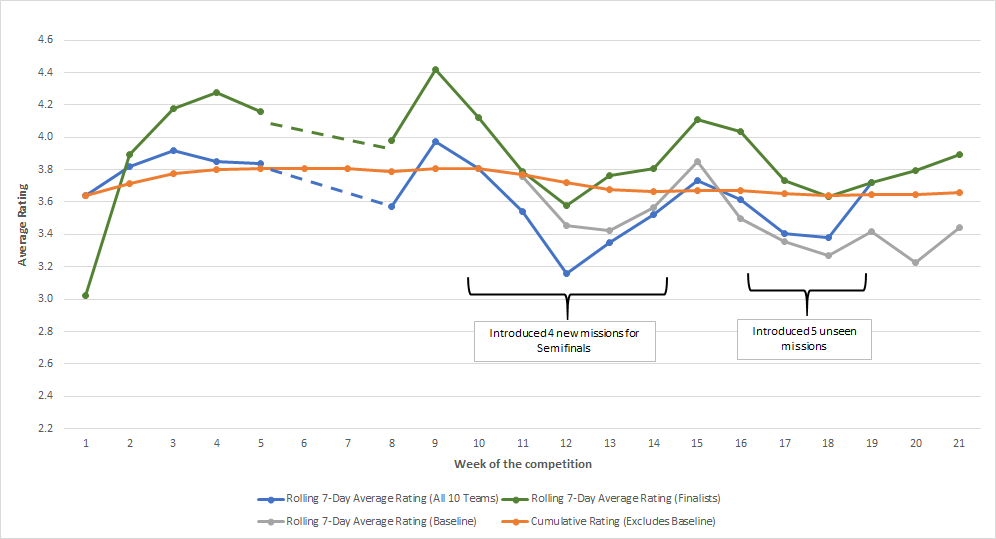}
\caption{Rolling 7-Day Average Rating of User Satisfaction over the period of the competition for all SimBots (Blue), Finalists (Green), the progression of the cumulative ratings for all SimBots excluding Baseline (Orange), and the Baseline (Gray). The dashed green and blue line indicate weeks with missing data.}
\label{fig:l7d-ratings}
\end{figure}

The primary mechanism of evaluating the SimBots was capture of explicit user satisfaction ratings. After each interaction, Alexa users were asked to rate their interaction with the SimBot on a scale of 1-5, according to the prompt, "How would you rate your interaction with the robot?". It's important to note that the SimBot rating prompt differed from the prompt used in the SocialBot competition ("How do you feel about speaking with this SocialBot again?") and the Taskbot competition ("How helpful was this TaskBot in assisting you?"), and thus, the ratings should not be directly compared between the different competitions. As shown in Figure~\ref{fig:l7d-ratings}, the Finalists improved their rolling seven-day average ratings by $30\%$ (from 3.0 to 3.9) over the span of 22 weeks in the competition. The cumulative average rating across all teams also experienced an increase of $3\%$, progressing from 3.6 to 3.7 throughout the competition.

\subsection{Task Completion Metrics}

\begin{figure}[!h]
\centering
\includegraphics[width=1\textwidth]{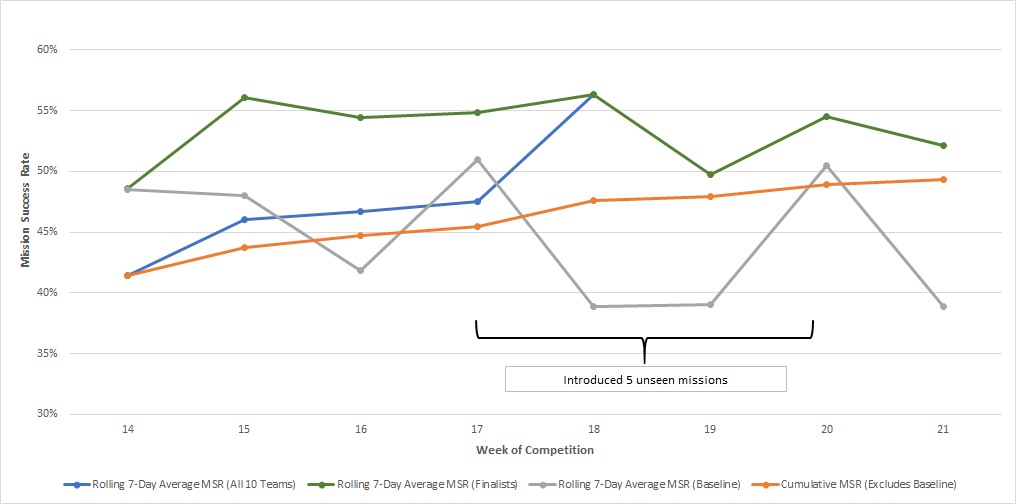}
\caption{Rolling 7-Day Average MSR over the period of the competition for all SimBots (Blue), Finalists (Green), the progression of MSR for all SimBots excluding Baseline (Orange), and the Baseline (Gray).}
\label{fig:l7d-msr}
\end{figure}

In addition to the satisfaction ratings, the Mission Success Rate (MSR) was introduced as a task oriented metric in Week 13. The mission success rate for each team was calculated by dividing the number of successful missions by the total number of missions played by that team. As shown in Figure~\ref{fig:l7d-msr}, the Finalists improved their rolling seven-day average MSR by $4\%$ (from $49\%$ to $52\%$) over 8 weeks of the competition. The cumulative average MSR across all teams also increased by $8\%$ during the course of the competition, from $41\%$ to $49\%$.

In Week 17 of the competition, we introduced five new game missions that the SimBots had not previously seen. To successfully complete the unseen missions, the SimBots had to complete new actions and interact with new objects. Table~\ref{msr-table} presents the results comparing the seen and unseen missions. On unseen missions, the MSR for Finalist teams improved by $2\%$ from $53\%$ to $55\%$ and all 10 university teams improved by $2\%$ from $45\%$ to $47\%$. The baseline system exhibited an improvement of $10\%$ from $45\%$ to $55\%$ on the unseen missions.

Notably, a high correlation between customer satisfaction (CSAT) and MSR was observed across all teams. During the Semifinals, there was a correlation of 0.92 (Pearson’s Correlation) between CSAT and MSR across all 10 university teams, highlighting the strong relationship between user satisfaction and task completion.

\begin{table}[!h]
\begin{center}
\begin{tabular}{|l|c|c|c|}
\hline
\textbf{MSR}             & {\textbf{Seen Missions}} & \textbf{Unseen Missions} & \textbf{Variance} \\ \hline
\textbf{All 10 teams}    & 45\%                                       & 47\%                     & 2\%               \\ \hline
\textbf{Finalist teams}  & 53\%                                       & 55\%                     & 2\%               \\ \hline
\textbf{Baseline System} & 45\%                                       & 55\%                     & 10\%              \\ \hline
\end{tabular}
\caption{\label{msr-table} MSR comparison for seen and unseen missions for all 10 teams, finalist teams, and our baseline system.}
\end{center}
\end{table}


\section{Discussion and Conclusions}
\label{sec:conclusions}



While substantial advances have been made in the application of AI to create compelling and useful conversational assistants, significant challenges remain in advancing from the digital domain to create embodied conversational agents that can navigate the real world, manipulate objects, and complete tasks. The SimBot Challenge enabled university teams from around the world to compete to create effective and usable embodied conversational AI agents that were able to operate in a simulated environment that was fielded to Alexa users. This was the first edition of the SimBot Challenge and developing a competition in embodied AI that could be used by Alexa users on multimodal devices was a very challenging mission. In addition to providing the infrastructure for creating robust interactive spoken dialog systems, we also had to design and build the Alexa Arena simulation environment, develop compelling game missions for everyday users, and ensure support for capturing the robot's first-person view and applying computer vision models. Teams ratings and mission completion rates improved steadily across the course of the competition and teams were able to create approaches that generalized to unseen objects and tasks.
The collaborative efforts of the Alexa Prize team and the participating university teams have laid a solid foundation for expanding the Alexa Prize SimBot Challenge, driving advancements in embodied conversational AI, and extending the possibilities for Alexa users in the future. 




\subsubsection*{Acknowledgments}

We would like to thank all the university students and their advisors (Alexa Prize SimBot Teams) who participated in the competition. We thank Amazon leadership and Alexa principals within the Alexa Natural Understanding (NU) organization for their vision and support through this entire program; Marketing for helping drive the right messaging and traffic to the Alexa Prize SimBot Challenge, ensuring that the participating teams received real-world feedback for their research; and Alexa Engineering for the work on enabling the Alexa Prize SimBot skills. We are grateful to the Alexa Developer Experience and Customer Trust (ADECT) Gadgets team for the many certification requests they worked on quickly to certify the university bots. We’d also like to thank the NU-Customer Experience team for exemplifying customer obsession by providing teams with critical inputs on building the best user experiences. We thank our leaders who took the time to virtually visit the university teams, learning from the teams and probing them to help them improve their designs. The competition would not have been possible without the support of all Alexa organizations including Speech, NLU, Data Services, and Conversation Modeling leadership. And finally, we would like to thank Alexa users who engaged in many interactions with the new Alexa Prize SimBot Challenge and provided feedback that helped teams improve over the course of the year.

\bibliography{main.bbl}

\begin{thebibliography}{10}

\bibitem{hu2021further}
S.~Hu, Y.~Liu, A.~Gottardi, B.~Hedayatnia, A.~Khatri, A.~Chadha, Q.~Chen,
  P.~Rajan, A.~Binici, V.~Somani, {\em et~al.}, ``Further advances in open
  domain dialog systems in the fourth alexa prize socialbot grand challenge,''
  2021.

\bibitem{gottardi2022alexa}
A.~Gottardi, O.~Ipek, G.~Castellucci, S.~Hu, L.~Vaz, Y.~Lu, A.~Khatri,
  A.~Chadha, D.~Zhang, S.~Sahai, {\em et~al.}, ``Alexa, let's work together:
  Introducing the first alexa prize taskbot challenge on conversational task
  assistance,'' {\em arXiv preprint arXiv:2209.06321}, 2022.

\bibitem{zobrist1969model}
A.~L. Zobrist, ``A model of visual organization for the game of go,'' in {\em
  Proceedings of the May 14-16, 1969, spring joint computer conference},
  pp.~103--112, 1969.

\bibitem{gao2023alexa}
Q.~Gao, G.~Thattai, X.~Gao, S.~Shakiah, S.~Pansare, V.~Sharma, G.~Sukhatme,
  H.~Shi, B.~Yang, D.~Zheng, {\em et~al.}, ``Alexa arena: A user-centric
  interactive platform for embodied ai,'' {\em arXiv preprint
  arXiv:2303.01586}, 2023.

\bibitem{teach}
A.~Padmakumar, J.~Thomason, A.~Shrivastava, P.~Lange, A.~Narayan-Chen,
  S.~Gella, R.~Piramuthu, G.~Tur, and D.~Hakkani-Tur, ``Teach: Task-driven
  embodied agents that chat,'' in {\em Proceedings of the AAAI Conference on
  Artificial Intelligence}, vol.~36, pp.~2017--2025, 2022.

\bibitem{ai2thor}
E.~Kolve, R.~Mottaghi, W.~Han, E.~VanderBilt, L.~Weihs, A.~Herrasti, D.~Gordon,
  Y.~Zhu, A.~Gupta, and A.~Farhadi, ``Ai2-thor: An interactive 3d environment
  for visual ai,'' {\em arXiv preprint arXiv:1712.05474}, 2017.

\bibitem{et}
A.~Pashevich, C.~Schmid, and C.~Sun, ``{Episodic Transformer for
  Vision-and-Language Navigation},'' in {\em International Conference on
  Computer Vision (ICCV)}, 2021.

\bibitem{vaswani2017attention}
A.~Vaswani, N.~Shazeer, N.~Parmar, J.~Uszkoreit, L.~Jones, A.~N. Gomez,
  {\L}.~Kaiser, and I.~Polosukhin, ``Attention is all you need,'' {\em Advances
  in neural information processing systems}, vol.~30, 2017.

\bibitem{cheng2021per}
B.~Cheng, A.~Schwing, and A.~Kirillov, ``Per-pixel classification is not all
  you need for semantic segmentation,'' {\em Advances in Neural Information
  Processing Systems}, vol.~34, pp.~17864--17875, 2021.

\bibitem{zhang2021vinvl}
P.~Zhang, X.~Li, X.~Hu, J.~Yang, L.~Zhang, L.~Wang, Y.~Choi, and J.~Gao,
  ``Vinvl: Revisiting visual representations in vision-language models,'' 2021.

\bibitem{hudson2019gqa}
D.~A. Hudson and C.~D. Manning, ``Gqa: A new dataset for real-world visual
  reasoning and compositional question answering,'' in {\em Proceedings of the
  IEEE/CVF conference on computer vision and pattern recognition},
  pp.~6700--6709, 2019.

\bibitem{lewis2019bart}
M.~Lewis, Y.~Liu, N.~Goyal, M.~Ghazvininejad, A.~Mohamed, O.~Levy, V.~Stoyanov,
  and L.~Zettlemoyer, ``Bart: Denoising sequence-to-sequence pre-training for
  natural language generation, translation, and comprehension,'' {\em arXiv
  preprint arXiv:1910.13461}, 2019.

\end{thebibliography}
\bibliographystyle{ieeetr}

\end{document}